# Remote Data Auditing and How it May Affect the Chain of Custody in a Cloud Environment

Rodolfo Machuca
Doctoral student (Cybersecurity)
Marymount University
Arlington, VA.USA
r0m54383@marymount.edu

Fatoumata Sankare
Doctoral student (Cybersecurity)
Marymount University
Arlington, VA. USA
f0s54147@marymount.edu

**Abstract**

As big data collection continues to grow, more and more organizations are relying on outsourcing their data to cloud-based environments. This includes the federal government and several agencies that depend on maintaining our citizens' secure data. Law enforcement agencies from the national level down to large city police departments are also using the cloud environment to store data. These agencies see this as a method of securing data while saving money by not maintaining the large data centers required to house this information. This data solution presents in own set of problems in that the outsourced data can become untrustworthy due to the lack of control of the data owners. Cloud computing is facing many difficulties, with security being the primary issue. This is because the cloud computing service provider is a separate entity; any data stored in the cloud can be interpreted as giving up control of the data by the primary data owner. [1] Remote data auditing (RDA) is increasingly important when managing data in a cloud environment, especially when organizations have to store their data in a multi-cloud environment. The challenging security threats posed by attempting to maintain the integrity of the data for proper auditing is a trial that was never addressed in the past. As the growth of cloud computing continues to dominate the digital information sphere, we need to address the demands of how to audit data correctly. To streamline the auditing of data, organizations are looking into RDA techniques. It's not to say that RDA techniques are the only viable or even suitable solution. The bulk of these techniques can only be applied to static archive data and not dynamically updated outsourced data. The data can be static, archival, backup, or dynamic. The data format must be able to support operations like insertion, deletion, and modification. One of the more challenging aspects is providing integrity for dynamic data, which is more challenging than static data or simply attaching data. RDA schemes are touted as the end-all-be-all solutions. The reality is that most RDA schemes cannot handle dynamic data characteristics or the demands that data integrity remains intact, given the opportunity for the data owner to insert, delete, or modify that data. [2, 3] We can now begin to see some of the limitations of RDA, and while there are many approaches to RDA, this paper will provide an overall view of the issues regarding remote data auditing while providing possible solutions for the future of RDA.

**Keywords**

Remote Data Auditing, Provable Data Possession, Multiple Replica, Dynamic Multi-Replica Data Possession

## I. INTRODUCTION

With the cost of procurement and the maintenance of large data centers, organizations are looking for ways to reduce the cost while getting an improved return on their investment. Enter cloud computing but more specifically, cloud storage. The ability to reduce the cost of storing data in a secure environment that provides a reliable and resilient infrastructure has afforded many organizations to adopt the notion of remotely storing their data while using on-demand applications to access that data from anywhere in the world. However, not everything that shines is gold. With that, the notion of storing information remotely produces a specific set of problems, with the most severe being how the integrity of the outsourced data is maintained. The loss of an organization to physically control its information is lost when you upload your data to a cloud data storage provider.

Therefore, organizations have found themselves designing Remote Data Auditing techniques that target security threats that target the outsourced data. When an organization implements RDA, it does so to allow its staff the ability to check the integrity of the stored data. Recently designed RDA approaches are categorized into three classes: replication-based, network coding-based, and erasure coding based, which present a taxonomy of the technique. These techniques are implemented to allow any organization to prove its reliability, which is essential in maintaining the security and validity of the data they have essentially relinquished control of.

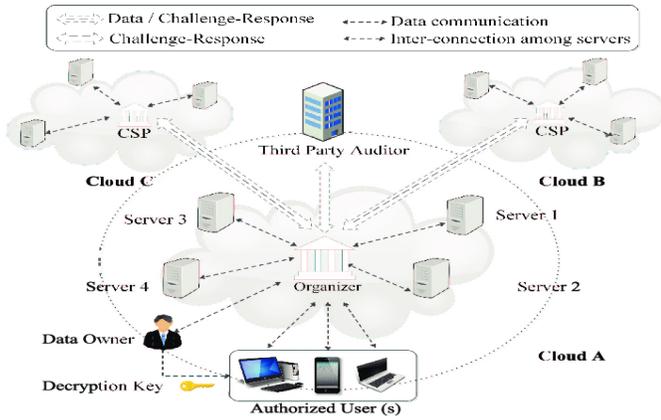

Fig. 1 Multi-cloud and Multi-server audit architecture

## II. LITERATURE REVIEW

Any Cloud computing offers organizations the option of not having to host or maintain data and data centers. For most companies, these are luxuries that place an incredible amount of financial strain on the purpose of keeping data secure for retrieval by its employees. Could services rely on pricing models based on usage and the amount of data an organization needs to store securely? Unlike traditional computing services, cloud computing utilizes virtualization to optimum use of the computing power available. With that said, could computing can be separated into three models: Software as a Service (SaaS), Infrastructure as a Service (IaaS), and Platform as a Service (PaaS). Each of these models provides different abilities to the organizations that implement them in the cloud environment. For example, IaaS allows organizations to establish processing, storage, and other essential resources. This includes the ability to control operating systems and other deployed applications. PaaS gives organizations the ability to deploy directly onto the cloud infrastructure. However, the organization will not be able to control the cloud infrastructure, which includes the servers or storage. An organization's only control with PaaS is the ability to manipulate deployed applications. Lastly, SaaS allows any organization that implements this service to use the various applications operating on a cloud infrastructure. SaaS limits the organization by not allowing them to manage the cloud infrastructure, which includes network servers or individual applications.

## III. METHODOLOGY

In its simplest terms, the chain of custody is the ability to provide detailed information that can track a record that began with its collection from the point it was either collected or received. The collection process is the root of any investigation and could be the most crucial step in documenting the collected evidence. It is vital to accurately record the collection of any evidence to validate the findings. This is especially important when the evidence is going to be used to go to trial, which could last months or even years. If the chain of custody for the evidence were done correctly at the onset of the investigation, it would be easier to locate the necessary information and prove that it has not been altered in any manner.

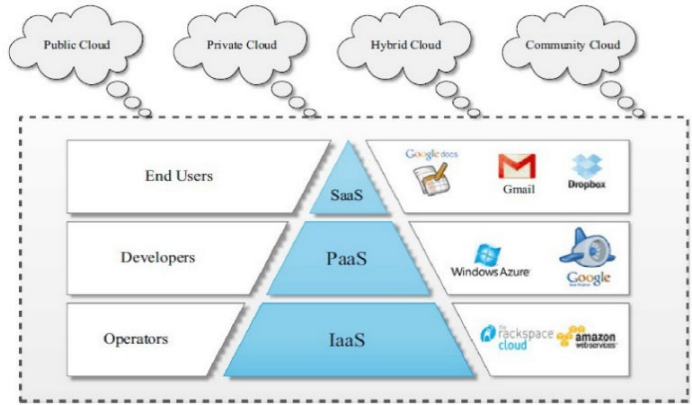

Fig. 2. Different service delivery models in cloud computing

This is a key aspect of manipulating evidence within the chain of custody because any data which may seem as if it was tampered with will be inadmissible in court. The agency bringing the suit against the defendant must be able to prove beyond a doubt that the collection of the evidence and subsequent storage was free of tampering. The use of the cloud environment to preserve digital data, which is becoming more and more prevalent in trial courts, is coming into question when addressing the chain of custody and whether the stored data is manipulated in any manner by the third-party vendor the data.

The argument can be made that the data is out of the hands of the organization storing it. The data is possibly being stored thousands of miles away, and the responsibility of how the data is secured is in the hands of the data storage company. Another question regarding the chain of custody is whether the data is being sent automatically from one server to another. Or is there a person or group who must first do something to the data before it is uploaded? While data storage in the cloud may seem secure, it is not infallible. The risk of cyber threat actors breaching these cloud storage systems is a reality.

Additionally, how can cloud data storage providers guarantee the privacy of data? Irrespective of the data being breached by a hacker, the data stored on these cloud storage systems can be viewed by personnel of these organizations; it is up to the third-party data storage organization to deny access to uploaded data to their personnel. Lastly, replicating data from one cloud environment to another poses yet another issue when asserting that the chain of custody of data has not been violated. The replication of data, whether for server maintenance or because the data has outgrown its current storage capacity, can be seen as manipulation of the data and an infringement of the chain of custody.

One of the most significant challenges of replicating data across several cloud storage environments is the possibility of service interruptions while the data is transferred from one server to another. Also, maintaining the data currently can be difficult due to the locations where the data may be stored. This results in data storage organizations having to implement complex data tracking systems to keep accurate track of data. This leads to remote data auditing as a possible solution to maintaining the validity of data stored in the cloud.

## IV. REPLICATION-BASED REMOTE DATA AUDITING

Any organization that stores its data in a cloud environment must contend with several issues, one of which is data failures. This is when redundancy plays a pivotal role in retrieving outsourced data. Implementing replication techniques offers a solution in that multiple copies of the data are stored within a distributed storage system. This allows the organization to retrieve uncorrupted data if one of the copies were ever to become corrupted. This solution does have its drawbacks in that having multiple copies of the same data occupies more space on the cloud servers, and this space is not inexpensive.

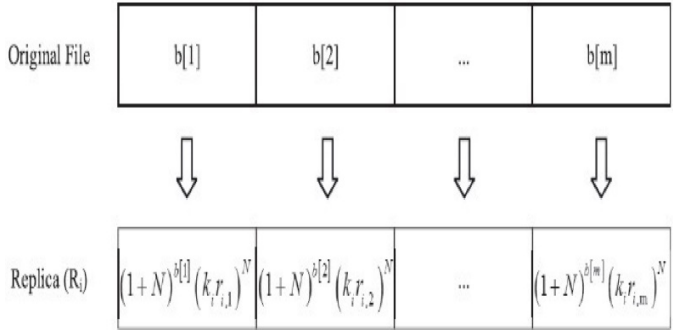

Fig. 3. Generating a unique replication in the DMR-PDP scheme.

## V. ERASURE CODING BASED REMOTE DATA AUDITING

Another class of RDA is the erasure code technique which implements a data repairing technique. Unlike the replication technique, the erasure code technique provides the organization with more reliability of the stored data while allowing for the same redundancy as the replication-based technique. The erasure code allows the client to take advantage of implementing intact blocks to cipher the codes of the corrupted blocks. Just like the replication technique, the erasure code does have a drawback in that the communication overhead is substantially higher than the replication technique.

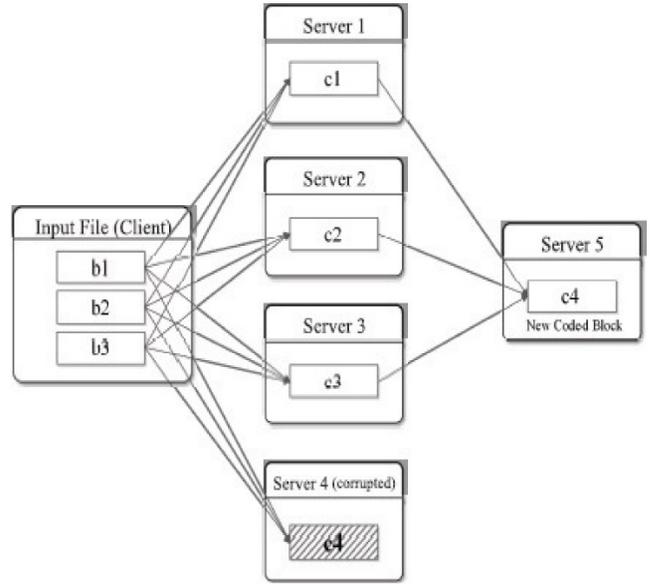

Fig. 4. Erasure-coding-based distributed storage system.

## VI. NETWORK CODING BASED REMOTE DATA AUDITING

This technique addresses the communication overhead issue with the erasure coding technique during the repair process. Instead of recomputing the data blocks like the previous technique, network coding detects the corrupt data block and creates a new data block based on the linear combination of the stored blocks over several servers.

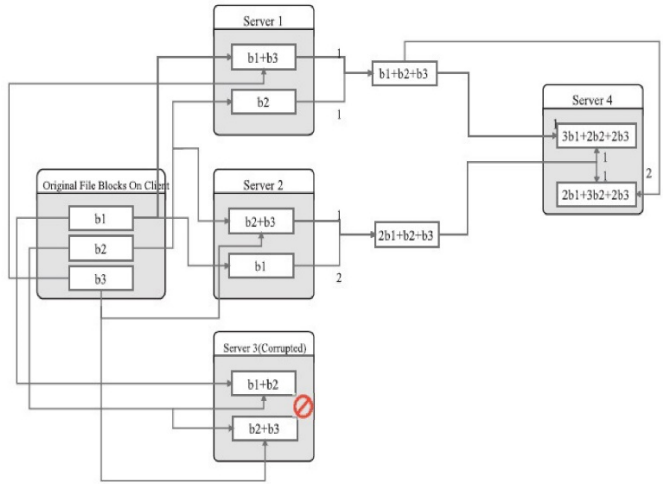

Fig. 5. Network-coding-based distributed storage system.

## VII. FINDINGS

The issue researchers have encountered when proposing remote data auditing protocols is how to accurately and efficiently verify the integrity of data uploaded and stored in the cloud. Remote data auditing is divided into integrity-based, recovery-based, and de-duplication-based categories. Each of these categories brings its level of responsibility to the data owner and the cloud-based data solution provider. Integrity-based: it enables a cloud user to verify the integrity of data, recovery based: data recovery is performed by leveraging error correction and erasure codes; however, normal integrity verification provides a way for recovering data in case of any possible corruption, and deduplication-based: it is meant to improve the efficiency of data storage and mitigate the communication overhead of data outsourcing.[4] Integrity-based: enables the cloud user to validate the integrity of data, recovery based: data recovery is achieved by leveraging fault corrections and deletion codes; however, standard integrity confirmation offers a way for recovering data in case of possible corruption, and deduplication-based: is intended to advance the effectiveness of data storage and mitigate the communication of data outsourcing. [5] Therefore, special attention must be paid when developing a suitable remote data auditing process.

The steps necessary for the RDA to be implemented are efficiency and a mode of verifiability. Additionally, frequency, probability of detection, and finally, dynamic update. Efficiency is needed to verify the data as quickly and as accurately as possible, while verifiability is a dual-fold mechanism. If private verification is required for a client's computer, verifying the data will differ from if a public verification method needs to be used. In that case, a third-party auditor or TPA is used to verify. The reasoning behind using a TPA is an attempt to apply considerably more processing power to the verification process of public data, which would not be needed in a private computer setting. When dealing with the frequency, we address the number of times a client can and should access and verify the data. This leads us to the probability of detection, which will give the data owner the likelihood of how much data loss is discovered while auditing is undertaken. Finally, we address the dynamic update, which will allow the data owner to authenticate the integrity of the data without downloading any of the data. This is important because any data capable of being downloaded is also susceptible to modification, including but not limited to insertion, deletion, and modification. The focal point of the project work is used to validate that the effort delivered an outcome that solves the problem statement.

## VIII. DISCUSSION

When addressing the problem of accurately auditing data, several solutions can be implemented; one such solution is implementing an efficient remote data auditing method based on data replication. Due to the large amount of data that is being collected, organized, and analyzed. Implementing this method will allow the data owner to check the data in the cloud environment efficiently, allowing for fewer processes to be implemented on the cloud and client-side compared to other cryptosystems. Additionally, the data auditing system can encompass an effective data structure that will support dynamic data updates. Implementing the use of this data structure will allow for dynamic data operations, such as the modification of specific processes that will enable this auditing method to be applied to large data sets while using the minimum amount of processing power on the client and server-side and allowing for multiple copies of the data to exist. This method will also consider the appropriate level of security that needs to be implemented to justify the data auditing method, given the possibility that there could be failures at several points of the collection or storage of data. An auditing plan that considers the ability to build in a redundancy offers a significant role in providing data reliability. When dealing with the potential loss of data due to any of the many failures that may occur, implementing a replication procedure is vital to the success of the audit. The data owner can use an initial complete copy of the file with size $|f|$ from any of the $r$ servers. However, implementing the replication method to take up the data storage space is $r|f|$.[6] While many people think that data uploaded to the cloud is somehow replicated onto multiple servers is incorrect. Another way to look at this is that in the replication-based data storage systems, we will not know whether our data housed in the server has multiple backups. [7] For example, in peer-to-peer networks, servers will perform a freeloading attack to use excessively more of the system's resources without contributing a balanced amount of resources back to its peers [8]. This will result in the data owners experiencing a deterioration in the resiliency and accessibility of the file. This will cause the CSP to have more storage space to sell to the users. [9]. A proven method to avoid the problems mentioned above is to implement a single provable data possession (PDP) method. Where $t$ times for $t$ different servers to get passed the issues created by the situation of a peer-to-peer network and the decline in the durability and availability of the file.

Utilizing servers that can work in unison while simultaneously pretending that $t$ copies of the data are stored on the server when in reality, only one copy is stored as an identical copy of the data is stored on all of the $t$ servers. The issue with implementing this method is the cost of analyzing the data is too high, and it is not practical for distributed storage systems, especially when dealing with larger data sets. A probable solution to replicating data sets and auditing data that has been copied on multiple servers across a cloud environment is the implementation of multiple replicas provable data possession (MRPDP). This describes whoever is the first to tackle the collude attack in the replication-based schemes. In this scheme, the data owner encrypts the original file and hides the encrypted file's blocks through an arbitrary value (Ru). Each copy will generate $t$ unique replicas (Rs). Afterward, the data owner practice of Boneh-Lynn-Shacham (BLS) homomorphic linear authenticators. The scheme is comprised of two versions: Deterministic (DEMC-PDP) and Probabilistic (PEMC-PDP). The deterministic version showed the file blocks and the probabilistic interpretation informally reviews the file. The primary method of the scheme is to produce an exclusive replication of the file by attaching a replica identifier to the original file. The data owner can use the equation which will formulate a tag for each block of replicas and apportion the tag throughout different servers.

$T_{i,j}=(h(F_{id})ub[i][j])d,$

($F_{id}$ indicates an inimitable fingerprint for each file produced by attaching the filename to the identifier of blocks and the number of replicas. I indicate the replication number, J is the block number, D is the client's private key, and U is a generator for a bilinear group mapping G. Even though this scheme supports the reliable third-party auditing test to bring up-to-date a block of the file, the data owner must re-encrypt and then upload all replicas to the various servers. While this scheme offers security, it also produces more significant storage overhead at both the client and server ends. [11] Furthermore, Dynamic Multi-Replica Data Possession (DMR-PDP) has been introduced as a scheme to verify the integrity of multiple copies of the same data set. This scheme implements the Paillier encryption technique to create copies of the original file. There is also the use of a Homomorphic verifiable response to integrate several responses. This technique is also the primary way cooperative provable data possession works as it reduces the bandwidth while hiding the location of the data within the cloud environment. Ultimately, whichever solution or scheme is implemented, RDA aims to relieve the security threats outsourced data poses to the environment it resides in by checking the reliability and integrity of the data stored in a single server or distributed server environment.

## IX. CONCLUSION

Many organizations rely on the cloud to store data by many organizations, which has increased the need for Remote Data Auditing. Various techniques were introduced for RDA to be successful, and RDA was implemented to check the integrity of the stored data. Remote auditing is a vital part of a data store because of the need to maintain the data's security, validity, and integrity. The chain of custody is also critical because it will provide a track record for the data stored in the cloud. For example, in a judicial proceeding, it will be easier to locate the data you need and prove that it was not manipulated in any way. This will help eliminate evidence being inadmissible in court.

Moreover, remote data auditing is a possible solution to maintaining the validity of data stored in the cloud because maintaining the data can be complex depending on the locations of the storage. RDA is divided into three categories: integrity-bard, recovery-based, and de-duplication-based. Though RDA may raise concerns for some, we have provided a possible solution to include implementing a remote data auditing method based on data replication, dynamic data updates, and a replication procedure. RDA has many effects on the chain of custody, but the goal is to maintain the integrity and originality of the data.